# Apparent thermal conductivity governed by interface-injected phonon spectra

D. G. Jeong[1,2], Terumasa Tadano[3], and J. S. Lee[1,2,*]

[1]*Department of Physics and Photon Science, Gwangju Institute of Science and Technology (GIST), Gwangju 61005, Korea*

[2]*Department of Physics, Korea Advanced Institute of Science and Technology (KAIST), Daejeon 34141, Korea*

[3]*Research Center for Magnetic and Spintronic Materials, National Institute for Materials Science (NIMS), 1-2-1 Sengen, Tsukuba, Ibaraki 305-0047, Japan*

*Corresponding author: jslee1@kaist.ac.kr

Abstract:

Thermoreflectance techniques such as time- and frequency-domain thermoreflectance (TDTR and FDTR) are widely used to measure the thermal conductivity of bulk materials, thin films, and interfaces. When phonon mean free paths $\Lambda_\omega$ become comparable to the thermal penetration depth, a Fourier analysis yields an apparent thermal conductivity $\kappa_{app}$ that deviates from the bulk value $\kappa_{bulk}$. Recent studies have highlighted the role of the phonon spectrum injected through the transducer interface, which is generally out of equilibrium owing to frequency-dependent transmission, and have shown that accounting for this nonequilibrium distribution can yield $\kappa_{app}>\kappa_{bulk}$, unlike conventional suppression-function analyses that predict only $\kappa_{app}\leq\kappa_{bulk}$. However, the spectral characteristics governing whether $\kappa_{app}$ is enhanced or reduced remain unclear. Here, we resolve the surface-temperature response of semi-infinite Si mode by mode and show that short-$\Lambda_\omega$-weighted injection raises the surface temperature, whereas long-$\Lambda_\omega$-weighted injection lowers it. This behavior is governed by how deep the injected phonons travel before thermalizing with the lattice, which is set by their mean free paths, and by how effectively the substrate conducts the injected energy away, which is limited by its long-mean-free-path phonons. For an Al (70 nm)/Si structure, Fourier fits of synthetic TDTR phase signals give $\kappa_{app}$ of Si from about 93 to 360 W m$^{-1}$ K$^{-1}$ for different interfacial transmission spectra, and $\kappa_{app}$ correlates with the heat-flux-weighted $\Lambda_{\omega}$ of the transmitted phonons. Thus, $\kappa_{app}$ from thermoreflectance reflects not only the substrate thermal properties but also the phonon spectrum injected across the transducer interface.

Modern nanoelectronic and photonic devices contain increasingly complex material interfaces and reduced characteristic dimensions, making efficient heat dissipation a growing challenge [1-3]. Heat produced in active regions often traverses thin films and multiple interfaces before reaching a heat sink, where the relevant length scales can become comparable to the phonon mean free path ($\Lambda_\omega$). Under these conditions, phonon transport becomes quasi-ballistic and can no longer be fully described by the diffusive picture underlying Fourier's law [4-6]. Because phonons span a broad range of $\Lambda_\omega$, their contributions to heat transport depend strongly on the relevant transport length scale [7-9]. Therefore, a quantitative understanding of this $\Lambda_\omega$-dependent phonon transport is essential for predicting heat dissipation and thermal resistance in thin-film heterostructures and nanoscale devices.

Nanoscale thermal transport is routinely investigated by time-domain thermoreflectance (TDTR) and frequency-domain thermoreflectance (FDTR) [10,11]. Importantly, however, the apparent thermal conductivity $\kappa_{\text{app}}$ extracted from thermoreflectance measurements can deviate from the intrinsic bulk thermal conductivity $\kappa_{\text{bulk}}$ in either direction: it is often reduced [12-14], but values above $\kappa_{\text{bulk}}$ have also been predicted and reported when the injected phonon population is nonthermal [15,16], indicating that the measured thermal response does not necessarily reflect the intrinsic diffusive transport properties of the material. The thermal probing depth in thermoreflectance measurements is commonly characterized by the thermal penetration depth, $d_{\text{th}} = \sqrt{\kappa_{\text{bulk}}/\pi C_{\text{bulk}} f_{\text{th}}}$, with $\kappa_{\text{bulk}}$, $C_{\text{bulk}}$, and $f_{\text{th}}$ denoting the thermal conductivity, volumetric heat capacity, and heating frequency, respectively. This length scale is the characteristic distance over which periodic temperature oscillations penetrate into the sample. Phonons with $\Lambda_\omega > d_{\text{th}}$ contribute less to the measured thermal response, so that a Fourier analysis yields $\kappa_{\text{app}} < \kappa_{\text{bulk}}$ (Fig. 1(a)) [17]. This reduction is commonly described by a phonon suppression function based on the ratio of $\Lambda_\omega$ and $d_{\text{th}}$, and has been widely used to reconstruct $\Lambda_\omega$ spectra from thermoreflectance measurements [18-20]. Standard suppression-function analyses rely on equilibrium heat-source assumptions [17,19,21]. In a thermoreflectance measurement, however, frequency-dependent phonon transmission across a transducer/substrate interface causes the injected phonons to deviate from the equilibrium distribution at the substrate temperature [6,22]. Several studies have examined the consequences of this nonequilibrium injection. Wilson and Cahill attributed part of the deviation of $\kappa_{\text{app}}$ from $\kappa_{\text{bulk}}$ to nonequilibrium among phonon modes near the heated interface [6], and Hua and Minnich showed that the spectral distribution of the heat source affects quasi-ballistic heat dissipation [23]. Hua *et al.* further showed that retaining the inhomogeneous source term, which had previously been neglected, allows $\kappa_{\text{app}}$ to exceed $\kappa_{\text{bulk}}$ as well as fall below it [16], and that the apparent agreement between $\kappa_{\text{app}}$ and $\kappa_{\text{bulk}}$ for Al/Si is a coincidence arising from the transmission profile of the Al/Si interface [22]. Chiloyan *et al.* predicted that a heat source emitting mainly low-frequency phonons with long mean free paths can yield heat transport beyond bulk heat conduction in a transient thermal grating geometry [15]. In thermoreflectance, by contrast, the injected spectrum is set by phonon transmission across the transducer/substrate interface, and one important question remains open: which property of this injected spectrum determines whether, and by how much, $\kappa_{\text{app}}$ exceeds or falls below $\kappa_{\text{bulk}}$.

In this work, we employ the phonon Boltzmann transport equation (pBTE) to calculate temperature depth profiles in semi-infinite Si with varying spectral distributions of surface heat generation, which represents the spectrum of phonons injected across the transducer/substrate interface, and examine how this distribution influences $\kappa_{\text{app}}$. Compared with Fourier predictions, a heat-generation spectrum weighted toward longer-$\Lambda_\omega$ phonons lowers the surface temperature and extends the temperature profile deeper into the substrate, corresponding to a higher $\kappa_{\text{app}}$. Conversely, a spectrum weighted toward shorter-$\Lambda_\omega$ phonons raises the surface temperature and confines the temperature variation closer to the surface, resulting in a lower $\kappa_{\text{app}}$, as illustrated in Fig. 1(b). This behavior results from two physically distinct contributions: the depth to which the injected phonons carry their energy before thermalizing, which depends on the injected spectrum, and the reduced effective conductivity of the substrate over the thermal penetration depth, which is largely a substrate property. We further simulate TDTR phase signals for an Al/Si structure with systematically varied interfacial phonon transmission spectra and fit them using a conventional multilayer Fourier model with $\kappa_{\text{app}}$ as one of the free parameters. The extracted $\kappa_{\text{app}}$ varies substantially with the spectrum of phonons transmitted into the Si substrate, showing a positive correlation with their heat-flux-weighted average $\Lambda_\omega$. These results indicate that $\kappa_{app}$ obtained by thermoreflectance is not a property of the substrate alone; it also depends on the spectrum of phonons injected across the transducer/substrate interface.

We introduce a formalism to describe how the temperature distribution in a semi-infinite solid is governed by an oscillatory volumetric heat-generation spectrum induced at the surface (Fig. 2(a)). The governing equations follow the spectral pBTE formulation developed by Hua and co-workers [22,24]. We briefly outline them and extend the analysis to a broad range of phonon $\Lambda_\omega$ and heat-generation spectra. For a thermal oscillation frequency $f_{th}$ (angular frequency $\eta = 2\pi f_{th}$) and a spatial angular frequency $\xi$ along the depth, the Fourier-transformed pBTE yields the temperature rise as

$$\Delta\tilde{T}(\xi,\eta) = \frac{\int_0^{\omega_m} \pi S_\omega v_\omega \frac{1+i\eta\tau_\omega}{(\Lambda_\omega\xi)^2}\log\left[1+\left(\frac{\Lambda_\omega\xi}{1+i\eta\tau_\omega}\right)^2\right]d\omega}{\int_0^{\omega_m}\frac{C_\omega}{\tau_\omega}\left[1-\frac{1}{\Lambda_\omega\xi}\tan^{-1}\left(\frac{\Lambda_\omega\xi}{1+i\eta\tau_\omega}\right)\right]d\omega},$$

where $\omega_m$ is the upper limit of the phonon frequency spectrum, and $v_\omega$, $\tau_\omega$, and $C_\omega$ are mode-specific group velocity, relaxation time, and volumetric heat capacity, respectively. The bulk thermal conductivity and the volumetric heat capacity are given by $\kappa_{bulk} = \int_0^{\omega_m}\frac{1}{3}C_\omega v_\omega \Lambda_\omega d\omega$ and $C_{bulk} = \int_0^{\omega_m} C_\omega d\omega$, respectively. The quantity $S_\omega$ characterizes the spectral strength of the volumetric heat-generation at the surface, yielding a spectral heat generation rate $\frac{\tilde{Q}_\omega(\eta,\mu,\xi)}{4\pi} = \frac{1}{2}S_\omega v_\omega|\mu|$. Integrating over the solid angle, with direction cosine $\mu$, gives the surface heat flux $\pi S_\omega v_\omega$. With the standard inverse transform $\Delta T(z,\eta) = \frac{1}{2\pi}\int_{-\infty}^{\infty}\Delta\tilde{T}(\xi,\eta)e^{i\xi z}d\xi$ and a mirror-image source representing the specular surface of the semi-infinite substrate (a factor of 2), the depth-dependent temperature profile becomes

$$\Delta T(z,\eta) = 2\int_0^\infty \frac{\int_0^{\omega_m} S_\omega v_\omega \frac{1+i\eta\tau_\omega}{(\Lambda_\omega\xi)^2}\log\left[1+\left(\frac{\Lambda_\omega\xi}{1+i\eta\tau_\omega}\right)^2\right]d\omega}{\int_0^{\omega_m}\frac{C_\omega}{\tau_\omega}\left[1-\frac{1}{\Lambda_\omega\xi}\tan^{-1}\left(\frac{\Lambda_\omega\xi}{1+i\eta\tau_\omega}\right)\right]d\omega}\cos(\xi z)d\xi.$$

In the purely diffusive limit $\tau_\omega \to 0$, the expression reduces to the classical Fourier's law solution as

$$\Delta\tilde{T}^F(\xi,\eta) \simeq \frac{\int_0^{\omega_m} \pi S_\omega v_\omega d\omega}{i\eta C_{\text{bulk}} + \xi^2 \kappa_{\text{bulk}}}.$$

For given $C_{\text{bulk}}$ and $\kappa_{\text{bulk}}$, Fourier's law predicts a unique temperature profile whose magnitude scales with the total heat flux. In contrast, the spectral pBTE allows different depth-dependent temperature profiles under the same conditions, depending on the spectral distribution of surface heat-generation. Consequently, interpreting such a pBTE-derived temperature profile using Fourier's law can yield an apparent thermal conductivity $\kappa_{\text{app}}$ that differs from $\kappa_{\text{bulk}}$. This discrepancy arises because Fourier's law does not capture the nonlocal nature of phonon transport and the spectral dependence of the volumetric heat-generation term in the pBTE [16]. Building on the work of Hua and co-workers, we investigate how these effects manifest in the depth-dependent temperature profile and, ultimately, in the $\kappa_{\text{app}}$ extracted from thermoreflectance measurements.

We first examine how the depth-dependent temperature distribution within a semi-infinite sample differs among three representative surface heat-generation spectra having distinct spectral characteristics. An oscillatory surface heat-generation $Q_\omega(z=0)$ $(=S_\omega v_\omega)$ at frequency $f_{\text{th}}$ is applied at the surface ($z$=0) of semi-infinite Si, as shown in Fig. 2(a). Regarding the ω dependence of $Q_\omega$, Profile #1 is weighted toward high-frequency phonons with relatively short-$\Lambda_\omega$, Profile #2 represents an intermediate distribution, and Profile #3 is weighted toward low-frequency phonons with relatively long-$\Lambda_\omega$. All three spectra are normalized to have the same integrated intensity (Fig. 2(b)). As shown in Fig. 2(c), at $f_{\text{th}}$ = 0.10 MHz, the short-$\Lambda_\omega$ spectrum (Profile #1) elevates the surface temperature above the Fourier prediction ($\Delta T^F$), whereas the long-$\Lambda_\omega$ spectrum lowers it (Profile #3). For a semi-infinite solid under a periodic surface heat flux, the Fourier solution gives a surface temperature amplitude inversely proportional to the thermal effusivity $\sqrt{\kappa C}$. With $C$ fixed at $C_{\text{bulk}}$, a Fourier interpretation therefore yields $\kappa_{\text{app}}/\kappa_{\text{bulk}} \approx (\Delta T^F_{z=0}/\Delta T_{z=0})^2$, where $\Delta T_{z=0}$ and $\Delta T^F_{z=0}$ are the pBTE and Fourier surface temperature amplitudes. A surface temperature above the Fourier prediction (Profile #1) thus corresponds to $\kappa_{\text{app}}<\kappa_{\text{bulk}}$, and one below it (Profile #3) to $\kappa_{\text{app}}>\kappa_{\text{bulk}}$. These results demonstrate that the surface heat-generation spectrum $Q_\omega$ determines the temperature depth profiles through its $\Lambda_\omega$- dependent excitation of phonons, such that analysis of that profile using Fourier's law can yield an apparent substrate thermal conductivity that is either lower or higher than the bulk value.

To capture the dependence on $\Lambda_\omega$ more directly, we examine temperature profiles for monochromatic oscillatory surface heat-generation. The upper panel of Fig. 3(a) shows single-mode, unity-power surface heat-generation spectra arranged in order of increasing $\Lambda_\omega$, where $\Lambda_\omega$ spans a broad range from 10 nm to 100 μm. The lower panel shows the corresponding relationship between $\Lambda_\omega$ and relaxation time $\tau_\omega$, obtained from first-principles calculations of intrinsic three-phonon scattering in bulk Si [25] (computational details in Supplementary Information S3). Figures 3(b–d) display the corresponding depth-dependent temperature profiles at three different thermal oscillation frequencies $f_{\mathrm{th}}$, with each profile color-coded to match the corresponding phonon mode in Fig. 3(a). At $f_{\mathrm{th}}$ = 1 Hz, all modal temperature profiles coincide with the Fourier prediction $\Delta T^F$ because the thermal penetration depth $d_{\mathrm{th}}$ exceeds the $\Lambda_\omega$ of all phonon modes. Consequently, fitting these profiles with Fourier's law yields $\kappa_{\mathrm{app}} = \kappa_{\mathrm{bulk}}$, regardless of the heat-generation spectrum. Under such low-frequency excitation, heat transport therefore lies in the diffusive limit.

As the thermal oscillation frequency increases, the depth-dependent temperature profiles increasingly deviate from the Fourier solution. At $f_{\mathrm{th}}$ = 0.10 MHz, $d_{\mathrm{th}}$ is about 17 $\mu$m, and most phonon modes have $\Lambda_\omega$ shorter than $d_{\mathrm{th}}$, whereas some phonons in the low-frequency region have $\Lambda_\omega$ comparable to or longer than $d_{\mathrm{th}}$. Although the variation among temperature depth profiles generated by different $\Lambda_\omega$ phonons is relatively small, a systematic trend can be observed. Compared with $\Delta T^F$, profiles generated by short-$\Lambda_\omega$ phonons exhibit higher surface temperatures and tend to decay over a slightly shorter depth scale, reflecting a greater heat accumulation near the surface. Conversely, profiles generated by long-$\Lambda_\omega$ phonons exhibit lower surface temperatures and tend to extend slightly deeper, consistent with heat transport over greater depths (Supplementary Information S1). At $f_{\mathrm{th}}$ = 80.1 MHz, $d_{\mathrm{th}}$ is as small as 0.6 $\mu$m, and these deviations are further enhanced (Fig. 3(d)). In particular, the contrast between short- and long-$\Lambda_\omega$ phonons becomes more pronounced, with stronger near-surface heat accumulation for the former and deeper heat penetration for the latter. Accordingly, for the same injected energy, short-$\Lambda_\omega$ modes produce higher surface temperatures than long-$\Lambda_\omega$ modes, with the difference becoming more pronounced at higher thermal oscillation frequencies.

Importantly, although the oscillatory surface heat-generation is assigned to a single phonon mode, the resulting temperature profile reflects two contributions of the numerator and denominator in the expression of $\Delta T(z, \eta)$ that shift the surface temperature in opposite directions. The numerator incorporates the spectral weighting of injected phonons in determining the spatial temperature response. Because phonons can transport energy over distances related to their mean free paths, modes with longer $\Lambda_\omega$ can influence the temperature distribution at greater depths. Consistent with this picture, for modes with $\Lambda_\omega < d_{th}$ at intermediate $f_{th}$, the depth of maximum $|\nabla T|$ increases with $\Lambda_\omega$ (Supplementary Information S2). Where transport is locally diffusive, the heat flux is proportional to the temperature gradient; thus, while the depth of maximum $|\nabla T|$ indicates the characteristic depth to which the injected energy is carried, the actual energy release to the lattice occurs predominantly at shallower depths where the curvature $\partial^2 T/\partial z^2$ is large. Relative to the Fourier prediction, this lowers the surface temperature, with the effect becoming more pronounced as the injected spectrum shifts toward long-$\Lambda_\omega$ phonons. The denominator describes how effectively the substrate conducts this energy away from the surface. Substrate phonons with $\Lambda_\omega > d_{th}$ cross the steep near-surface temperature gradient with few collisions and contribute little to conduction across it; consequently, the effective conductivity over $d_{th}$ is lower than $\kappa_{bulk}$. This traps heat near the surface and raises the surface temperature largely independently of which modes are injected, analogous to the mechanism underlying $\kappa_{app} \leq \kappa_{bulk}$ for an equilibrium source [17,19]. It is most pronounced for the shortest-$\Lambda_\omega$ modes, whose energy is deposited near the surface, resulting in a surface temperature higher than the Fourier prediction (Fig. 3(e)). The balance between the two contributions determines whether $\kappa_{app}$ lies above or below $\kappa_{bulk}$.

We characterize the depth-dependent temperature profiles in terms of the surface temperature $\Delta T(z = 0)$ and the temperature decay depth $d_e$. The latter is defined as the depth at which the temperature falls to $1/e$ of its surface value, providing a characteristic length scale for an exponential-like temperature decay. Figures 3(e) and 3(f) show the normalized surface temperature $\Delta T(z = 0)/\Delta T^F(z = 0)$ and the normalized temperature decay depth $d_e/d_{th}$, respectively, over a wide range of thermal oscillation frequencies and phonon mean free paths (see Supplementary Information S1 and S2 for details). Two gray diagonal lines represent the $\Lambda_\omega$–$\tau_\omega$ relation shown in Fig. 3(a) for $f_{th}$ = 0.10 MHz and 50×80.1 MHz (≈4 GHz, the highest harmonic included in the conventional TDTR analysis). In the diffusive limit, all phonon modes satisfy $\Lambda_\omega/d_{th} < 1$, and both normalized quantities equal unity. As a fraction of modes enters the regime $\Lambda_\omega/d_{th} > 1$, heat transport crosses over to the quasi-ballistic regime, where short-$\Lambda_\omega$ modes exhibit $\Delta T(z = 0)/\Delta T^F(z = 0) > 1$ and $d_e/d_{th} < 1$, while long-$\Lambda_\omega$ modes produce the opposite trends. These deviations become more pronounced with increasing $f_{th}$ and reach their extremes in the ballistic limit, where all phonon modes satisfy $\Lambda_\omega/d_{th} > 1$ and $d_e/d_{th} < 1$. The analysis of individual phonon modes demonstrates that their contributions to the temperature distribution depend on both $\Lambda_\omega$ and the thermal oscillation frequency $f_{th}$. Consequently, individual modes exhibit different degrees of deviation from the Fourier prediction, even under identical $\kappa_{bulk}$ and $C_{bulk}$. This mode-dependent behavior provides the basis for extending the analysis to practical surface heat-generation spectra involving a broad range of phonon modes.

Since realistic surface heat-generation spectra span a broad range of phonon frequencies, we next consider weighted combinations of individual-mode contributions. Here, the weighting function $W$ scales as $\Lambda_\omega^1$, $\Lambda_\omega^0$, $\Lambda_\omega^{-1}$, corresponding to spectra dominated by long-$\Lambda_\omega$ phonons, equally contributed by all modes, and dominated by short-$\Lambda_\omega$ phonons, respectively (Fig. 3(g)). Because only the numerator of $\Delta T(z, \eta)$ depends on the spectral heat-generation $S_\omega$, $\Delta T(z, \eta)$ is linear in $S_\omega$, and the temperature profile for each weighted spectrum is given by the superposition of the single-mode temperature responses of Fig. 3, computed at the same $f_{th}$, scaled with the respective weights. As shown in Fig. 3(h), the normalized surface temperatures $\Delta T(z = 0)/\Delta T^F(z = 0)$ converge to unity in the low-frequency limit, regardless of the $\Lambda_\omega$ weighting, consistent with the Fourier prediction. As $f_{th}$ increases, however, the surface temperatures become increasingly dependent on the $\Lambda_\omega$ weighting; a spectrum weighted toward

long-$\Lambda_\omega$ phonons leads to lower surface temperatures, whereas greater contributions from short-$\Lambda_\omega$ phonons produce higher surface temperatures. These trends are consistent with the behavior observed in Fig. 2. Depending on the weight distribution $W$, $\Delta T(z = 0)/\Delta T^F(z = 0)$ exhibits a distinct functional dependence on $f_{th}$. Thus, the $f_{th}$ dependence of both $\Delta T(z = 0)/\Delta T^F(z = 0)$ and the corresponding $\kappa_{app}/\kappa_{bulk}$ obtained using a Fourier model reflects the spectral distribution of the surface heat generation.

We now consider thermal transport in a realistic TDTR geometry consisting of a 70-nm-thick Al transducer on a Si substrate. Here, phonons transmitted across the Al/Si interface define the surface heat-generation spectrum at the Si surface (Fig. 4(a)). We consider several types of transmission spectra having different spectral characteristics, which include the ones corresponding to Profiles #1, #2, and #3 shown in Fig. 2. For each transmission spectrum, the surface temperature is calculated using the model developed by Hua *et al.*, incorporating first-principles calculation results of phonon lifetimes, group velocities, and heat capacities of Al and Si [22,25] (Supplementary Information S4). TDTR phase signals are then constructed and subsequently fitted with the conventional multilayer Fourier heat-conduction model, treating both apparent thermal boundary conductance $G_{app}$ and apparent thermal conductivity $\kappa_{app}$ as free parameters (Supplementary Information S5). The construction of these transmission spectra and the extracted $\kappa_{app}$ as a function of the pump modulation frequency $f_{mod}$ are described in Supplementary Information S5. Profiles #1, #2, and #3 yield $\kappa_{app}$ values of about 93, 140, and 360 W m$^{-1}$ K$^{-1}$, respectively. In TDTR, the phase signal at the transducer surface is mainly sensitive to the interface conductance $G$ and the substrate effusivity. At the modulation frequencies used here ($f_{mod}$ = 5.40–14.5 MHz), $d_{th}$ in Si is about 2.3 μm–84 nm ($f_{th}$ = 5.40 MHz to ≈4 GHz). This length is comparable to the mean free paths of the Si phonons that carry a large fraction of the heat. In this regime, Fig. 3 shows mode-dependent deviations from the Fourier prediction, and the surface-temperature trends of Figs. 2 and 3 are therefore expected to be reflected in the fitted $\kappa_{app}$.

Importantly, the substrate $\kappa_{app}$ tends to increase with the heat-flux-weighted average $\Lambda_\omega$ of the phonons transmitted into Si (Fig. 4(b)). The heat-flux spectra are calculated with the same transmission spectra that are used to generate the TDTR phase signals. We consider a total of 505 spectra at $f_{th}$ =

$|f_{\text{mod}} \pm mf_{\text{rep}}|$ with $f_{\text{mod}}$ = 5.40, 7.10, 8.70, 10.0, 14.5 MHz, $m = -50, \ldots, 50$ and laser pulse repetition rate $f_{\text{rep}}$ = 80.1 MHz. The absolute value of the summed heat-flux spectra is then employed as the weighting factor in a weighted summation over the individual modal $\Lambda_\omega$. Here, $\langle \Lambda_\omega \rangle = \sum_\omega \Lambda_\omega |q_\omega| / \sum_\omega |q_\omega|$, where $q_\omega$ is the modal heat flux transmitted into Si, summed over all 505 heating frequencies. When short-$\Lambda_\omega$ modes dominate the transmitted heat flux (Profile #1), $\kappa_{\text{app}}$ falls below the bulk thermal conductivity of Si, whereas greater contributions from long-$\Lambda_\omega$ modes (Profile #3) lead to $\kappa_{\text{app}} > \kappa_{\text{bulk}}$. Several additional transmission spectra (Supplementary Information S5) follow the same trend. $\kappa_{\text{app}}/\kappa_{\text{bulk}}$ therefore correlates with the heat-flux-weighted mean free path $\langle \Lambda_\omega \rangle$ of the transmitted phonons. The transmission spectra considered here feature reduced high-frequency transmission while maintaining low-frequency transmission at the detailed-balance limit. Spectra with larger $\langle \Lambda_\omega \rangle$ therefore also tend to have smaller $G_{\text{app}}$ (Supplementary Information Fig. S5(b)), which lowers the sensitivity of the phase signal to $\kappa_{\text{app}}$ and strengthens the compensation between $G_{\text{app}}$ and $\kappa_{\text{app}}$ in the two-parameter Fourier fit; part of the spread in $\kappa_{\text{app}}$ may arise from this compensation. Nevertheless, the observation that short-$\Lambda_\omega$-weighted transmission yields lower $\kappa_{\text{app}}$ while long-$\Lambda_\omega$-weighted transmission gives higher $\kappa_{\text{app}}$ is consistent with the surface-temperature response of semi-infinite Si in Figs. 2 and 3. The distinct responses of individual phonon modes to varying thermal oscillation frequencies $f_{\text{th}}$ provide sensitivity to the spectral distribution of interfacial phonon transmission, and the fitted $G_{\text{app}}$ and $\kappa_{\text{app}}$ indeed depend on $f_{\text{mod}}$ differently for different transmission spectra (Supplementary Information Fig. S5(b, c)). This suggests that TDTR measurements over a range of $f_{\text{mod}}$ can provide information on the underlying interfacial transmission spectrum.

In conclusion, we have shown that the apparent thermal conductivity $\kappa_{\text{app}}$ obtained from a Fourier analysis of thermoreflectance data depends strongly on the spectrum of phonons injected across the transducer/substrate interface. Because phonon transmission across the interface depends strongly on frequency, the injected phonon population generally deviates from the equilibrium distribution at the substrate temperature. Incorporating this nonequilibrium distribution as a spectral heat-generation term in the phonon Boltzmann transport equation shows that it can lead to $\kappa_{\text{app}} > \kappa_{\text{bulk}}$. Nonequilibrium alone is not sufficient, however: injection weighted toward short-$\Lambda_\omega$ phonons raises the surface temperature above the Fourier prediction and yields $\kappa_{app} < \kappa_{bulk}$, whereas injection weighted toward long-$\Lambda_\omega$ phonons lowers it and yields $\kappa_{app} > \kappa_{bulk}$. This behavior results from the balance between how deep the injected phonons carry their energy, which depends on the injected spectrum, and the reduced effective conductivity of the substrate over the thermal penetration depth, which is largely a substrate property. Interfaces that transmit comparable total heat flux but with different spectral content are therefore expected to yield different $\kappa_{app}$ for the same substrate. A deviation of $\kappa_{app}$ from $\kappa_{bulk}$, in either direction, therefore need not indicate a measurement artifact or poor sample quality. It can instead reflect the phonon spectral selectivity of the transducer interface. Because measurements at different heating frequencies weight the injected phonon modes differently, modulation-frequency-dependent thermoreflectance data can contain information on the interfacial transmission spectrum [22]. This suggests a route toward probing, and ultimately engineering, spectral phonon transmission through interface roughness, interlayers, or defects.

**Acknowledgements**

D.G.J. and J.S.L. acknowledge support from the Korea government (MSIT) No. RS-2026-25549743 and RS-2026-25508298.

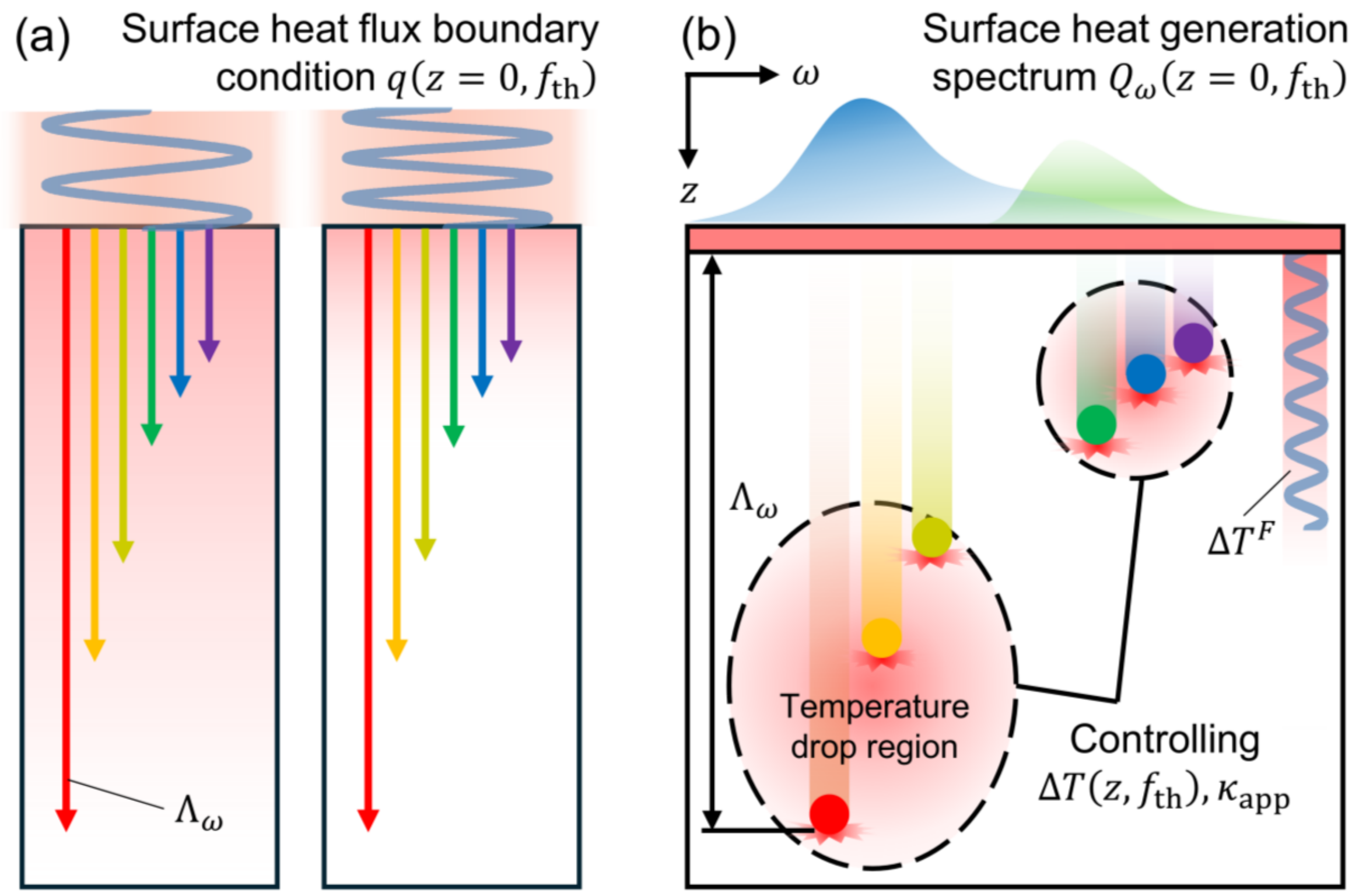

**Figure 1. Schematic illustration of thermal transport under oscillatory surface heating.** (a) Thermal transport in a semi-infinite material subject to an oscillatory surface heat flux. Colored arrows represent the phonon mean free paths ($\Lambda_\omega$) of individual modes, and the transparent red region illustrates the temperature distribution. As the oscillation frequency of the heat source increases, the thermal penetration depth decreases and becomes comparable to or shorter than $\Lambda_\omega$ of some phonons. (b) Thermal transport driven by spectrally distributed surface heat-generation, represented by the volumetric heat-generation term in the phonon transport equation. In the quasi-ballistic regime, the injected energy of each mode is carried to a characteristic depth that increases with $\Lambda_\omega$ before it is released to the lattice; for modes with $\Lambda_\omega$ shorter than the thermal penetration depth, the temperature gradient is largest near this depth, giving rise to deviations from Fourier's law predictions.

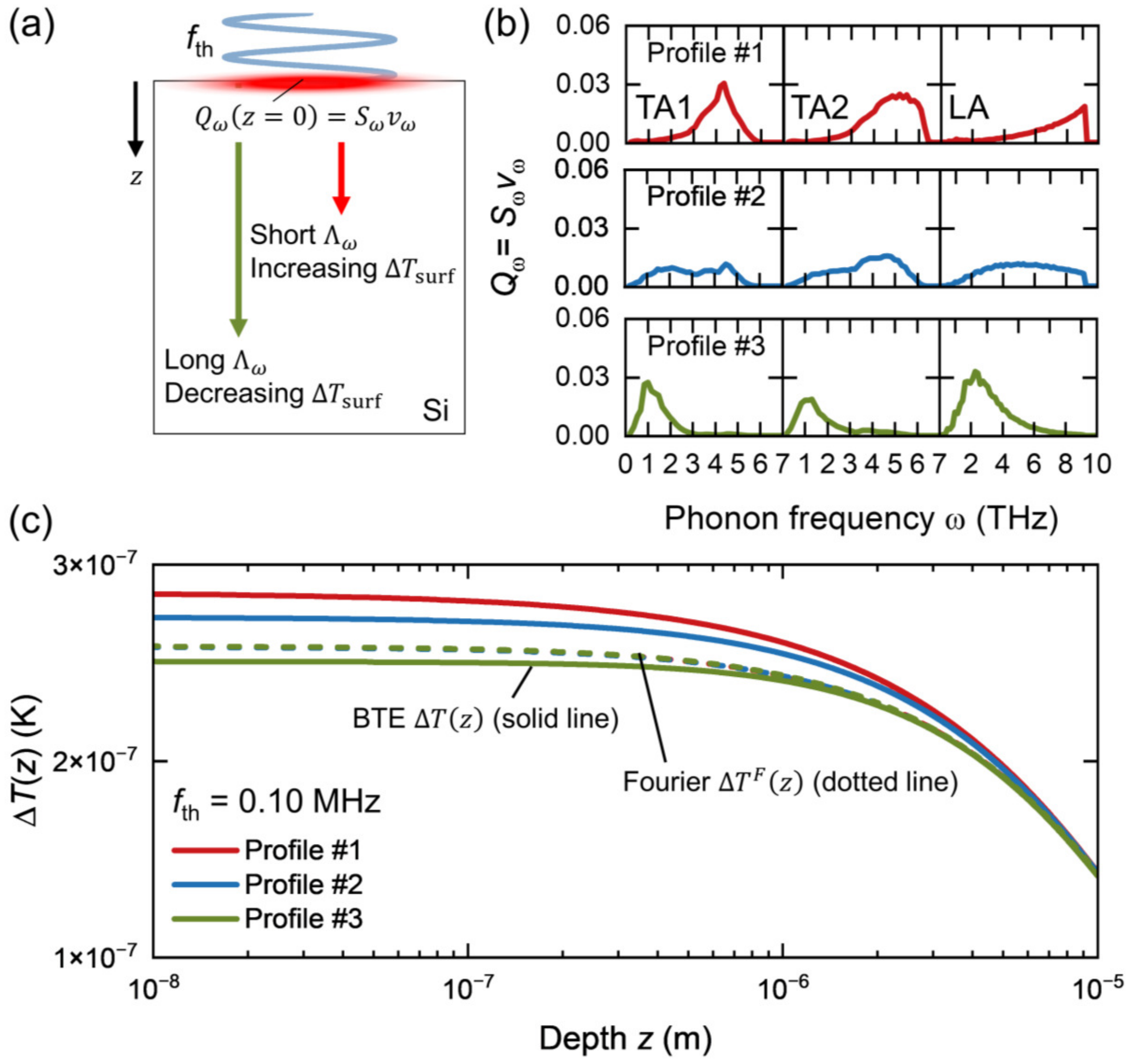

**Figure 2. Depth-dependent temperature profiles simulated using the phonon Boltzmann transport equation (pBTE) for different surface heat-generation spectra.** (a) Schematic of the simulation geometry. Oscillatory surface heat-generation $Q_\omega$ is applied at the surface of a semi-infinite silicon domain. (b) Three representative surface heat-generation spectra $Q_\omega$, labeled as Profiles #1–#3. (c) Corresponding temperature profiles as a function of depth for Profiles #1–#3 with thermal oscillation frequency $f_{th}$ = 0.10 MHz.

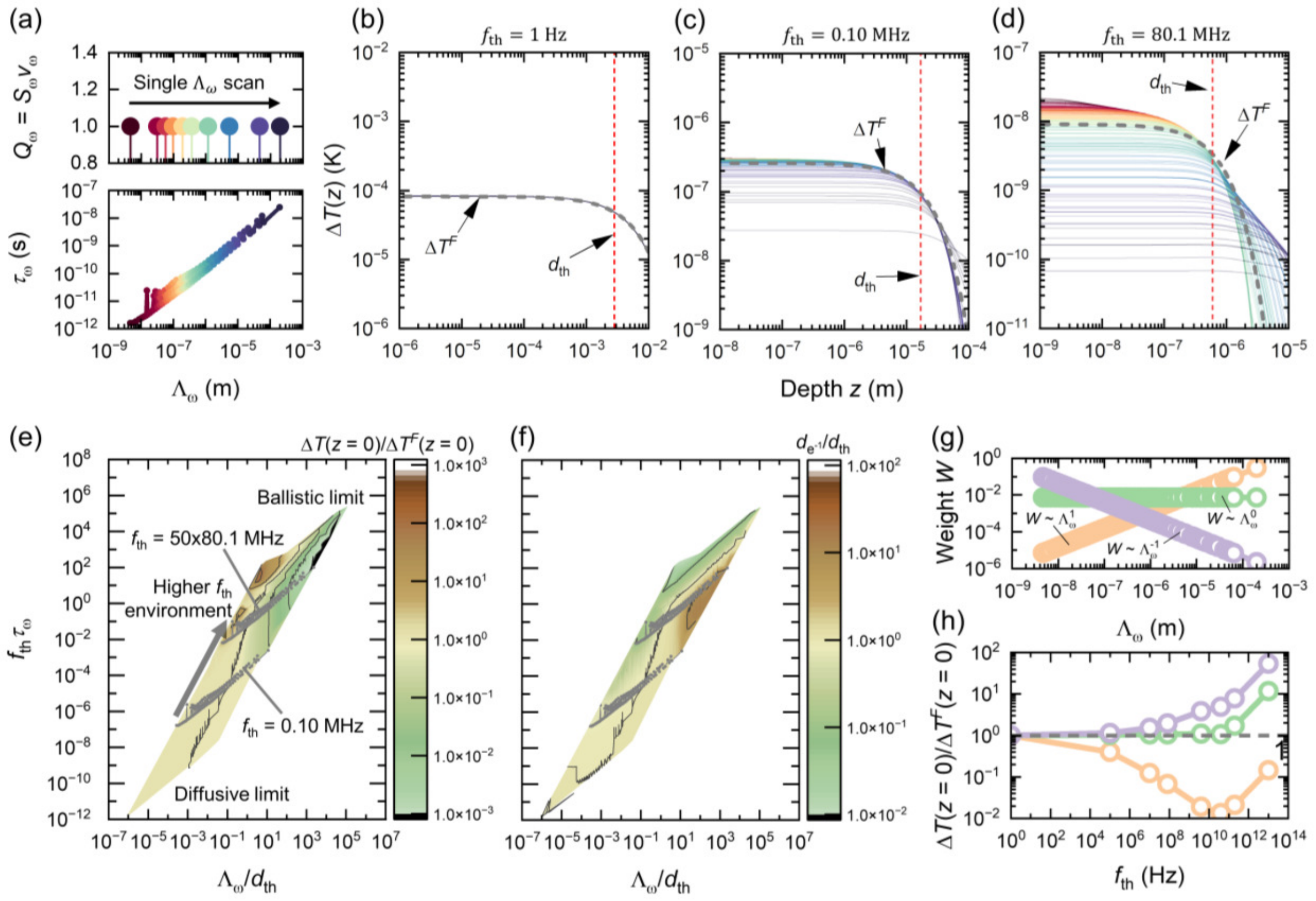

**Figure 3. Temperature responses under single-mode oscillatory surface heat-generation.** (a) Mean free path ($\Lambda_\omega$) spectrum of a single phonon mode and its relation to the relaxation time. (b–d) Depth-dependent temperature profiles under oscillatory surface heat-generation at frequencies $f_{th}$ of 1 Hz, 0.10 MHz, and 80.1 MHz, respectively. The vertical red dashed line indicates the thermal penetration depth $d_{th}$ calculated from Fourier's law. $\Delta T^F$ denotes the temperature depth profile predicted by Fourier's law. (e) Map of the surface temperature normalized by the Fourier prediction, plotted as a function of $f_{th}\tau_\omega$ and $\Lambda_\omega/d_{th}$. (f) Map of the temperature decay depth $d_e$ normalized by the Fourier prediction, plotted as a function of $f_{th}\tau_\omega$ and $\Lambda_\omega/d_{th}$. (g) Spectral weighting functions for the surface temperature as a function of $\Lambda_\omega$. (h) Normalized surface temperature obtained by applying the weighting functions in (g) to the single-mode responses shown in (e).

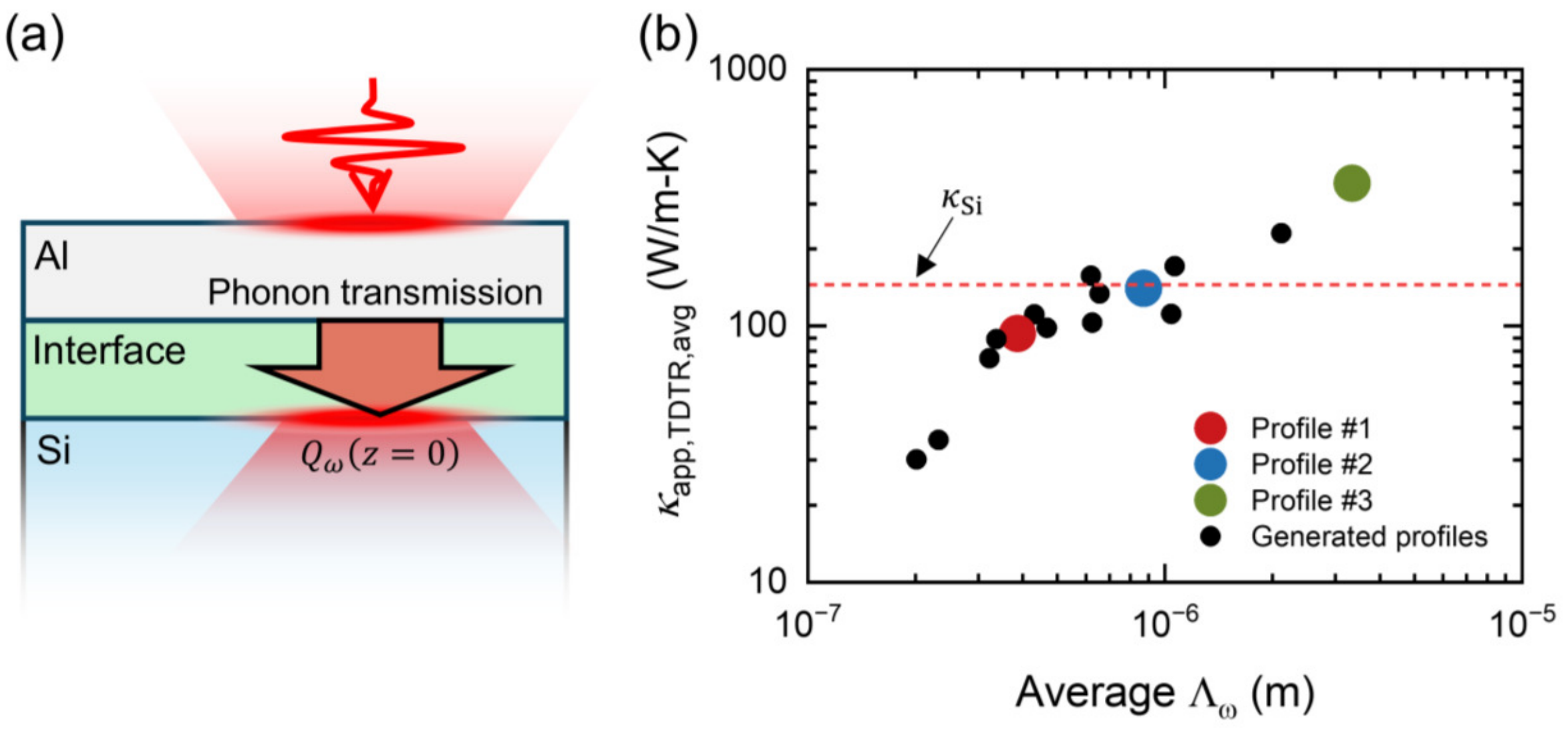

**Figure 4. Correlation between apparent substrate thermal conductivity determined from simulated TDTR phase signals and the average phonon mean free path.** (a) Geometry of the TDTR simulation model with a 70-nm-thick Al transducer on a Si substrate. Phonons transmitted across the Al/Si interface define the surface heat-generation spectrum in Si. (b) Relationship between the average apparent thermal conductivity and the average $\Lambda_\omega$. The apparent substrate thermal conductivity is obtained by fitting the pBTE-generated TDTR phase signals with the Fourier model at modulation frequencies $f_{mod}$ = 5.40, 7.10, 8.70, 10.0, 14.5 MHz, and the resulting values are averaged over these frequencies.

# Supplementary Information

# Apparent thermal conductivity governed by interface-injected phonon spectra

D. G. Jeong[1,2], Terumasa Tadano[3], and J. S. Lee[1,2,*]

[1]*Department of Physics and Photon Science, Gwangju Institute of Science and Technology (GIST), Gwangju 61005, Korea*

[2]*Department of Physics, Korea Advanced Institute of Science and Technology (KAIST), Daejeon 34141, Korea*

[3]*Research Center for Magnetic and Spintronic Materials, National Institute for Materials Science (NIMS), 1-2-1 Sengen, Tsukuba, Ibaraki 305-0047, Japan*

*Corresponding author: jslee1@kaist.ac.kr

Table of contents

S1. Temperature depth profiles normalized to the surface temperature for a single mode excitation

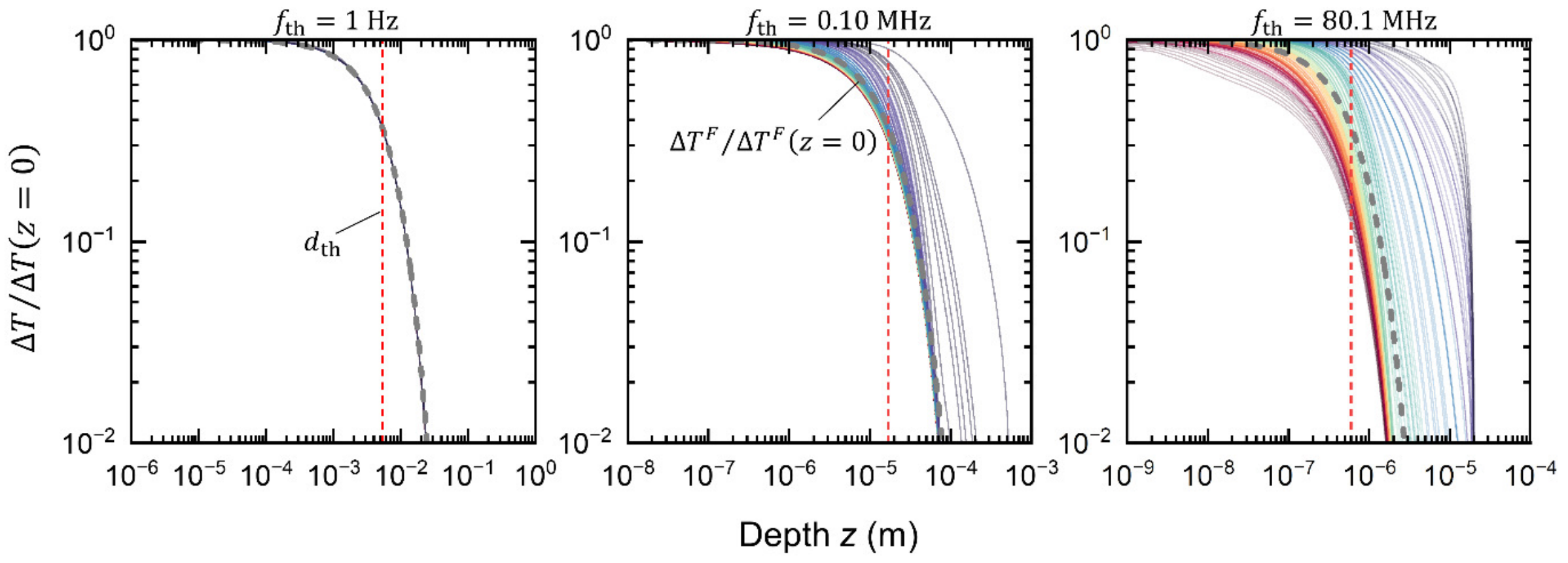

Figure S1. Temperature depth profiles normalized by the surface temperature of each mode. The vertical red dashed line indicates the thermal penetration depth obtained from Fourier's law. Line colors correspond to the same phonon modes as defined in Fig. 3(a) of the main text.

Figure S1 shows how the temperature decay depth $d_e$ of each mode compares with the Fourier prediction by normalizing the temperature depth profiles to the surface temperature of each mode. When the thermal oscillation frequency $f_{th}$ approaches zero, the depth profiles are indistinguishable from the Fourier result, corresponding to the diffusive limit. At higher thermal oscillation frequencies in the quasi-ballistic regime, the profiles deviate from the Fourier prediction; short-$\Lambda_\omega$ modes exhibit shorter temperature decay depths $d_e$, whereas long-$\Lambda_\omega$ modes exhibit longer $d_e$ than the Fourier prediction.

S2. Temperature gradient of single-mode depth profiles at various $f_{\text{th}}$

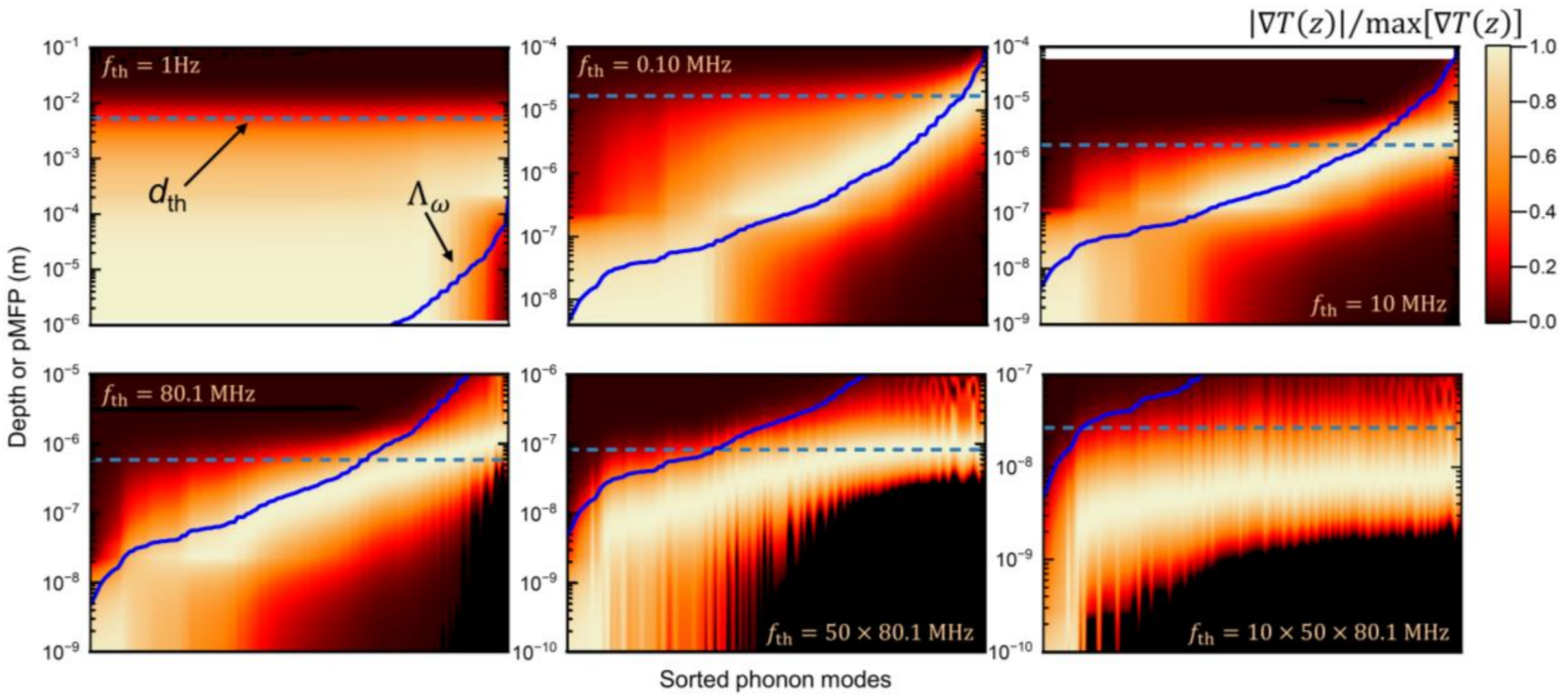

Figure S2. Normalized temperature gradient $|\nabla T(z)|/\max|\nabla T(z)|$ of single-mode temperature depth profiles at various thermal oscillation frequencies $f_{\text{th}}$; each phonon mode (column) is normalized to its maximum over depth, and the modes are sorted from short to long $\Lambda_\omega$. The blue line shows $\Lambda_\omega$ of each mode, and the horizontal dashed line indicates the thermal penetration depth determined by Fourier's law, $d_{\text{th}} = \sqrt{\kappa_{\text{bulk}}/\pi C_{\text{bulk}} f_{\text{th}}}$. For modes with $\Lambda_\omega < d_{\text{th}}$ at intermediate $f_{\text{th}}$, the maximum gradient lies near or slightly shallower than $\Lambda_\omega$; for modes with $\Lambda_\omega > d_{\text{th}}$, it remains near or shallower than $d_{\text{th}}$.

Figure S2 compares the depth of the maximum temperature gradient $|\nabla T|$ of each mode with its mean free path $\Lambda_\omega$ at various thermal oscillation frequencies $f_{\text{th}}$. At $f_{\text{th}}$ = 1 Hz, in the diffusive limit described in the main text, $\Lambda_\omega$ is shorter than $d_{\text{th}}$ for all modes, and the gradient profile is broad and similar for all modes, as in the Fourier prediction. At $f_{\text{th}}$ = 0.10 MHz and above, some phonon modes have $\Lambda_\omega$ longer than $d_{\text{th}}$, and thermal transport becomes quasi-ballistic. At $f_{\text{th}}$ = 0.10 MHz and 10 MHz, the depth of maximum $|\nabla T|$ increases with $\Lambda_\omega$ and lies near or slightly shallower than $\Lambda_\omega$ for modes with $\Lambda_\omega < d_{\text{th}}$, whereas it remains near $d_{\text{th}}$ for modes with $\Lambda_\omega > d_{\text{th}}$. At $f_{\text{th}}$ = 80.1 MHz, the maximum lies noticeably shallower than $\Lambda_\omega$, and at GHz frequencies it becomes nearly independent of $\Lambda_\omega$. Thus, short-$\Lambda_\omega$ modes carry their energy only to shallow depths, whereas long-$\Lambda_\omega$ modes carry it deeper, up to about $d_{th}$. Where heat transport is locally diffusive, the heat flux is proportional to $|\nabla T|$, so the maximum of $|\nabla T|$ marks the depth to which most of the injected energy has been carried; the energy is released to the lattice at shallower depths, where the curvature of the profile is large. This depth dependence is the numerator contribution in the expression of $\Delta T(z, \eta)$ discussed in the main text .

S3. Details of first-principles phonon calculations

The phonon properties of Al and Si used in this work, namely the phonon dispersions, mode heat capacities, group velocities, and intrinsic three-phonon relaxation times, were obtained from first principles using the ALAMODE package [1]. Harmonic and third-order interatomic force constants (IFCs) were computed in a 3×3×3 conventional supercell (216 atoms for Si, 108 atoms for Al). Harmonic IFCs were obtained from density functional theory (DFT) forces for a single 0.01 Å displacement and fitted by ordinary least squares. For the anharmonic IFCs, 20 randomly displaced configurations at 300 K were generated from the harmonic frequencies and polarization vectors, and IFCs up to sixth order within the fifth- (fourth-) nearest-neighbor shells of Si (Al) were extracted by adaptive LASSO (with NBODY=2 3 3 2 2 of ALAMODE). The three-phonon scattering rates were then computed on a dense $q$-point grid (100×100×100 for Si and 135×135×135 for Al). Within the single-mode relaxation-time approximation (RTA), the resulting room-temperature lattice thermal conductivities are 149 W $m^{-1}$ $K^{-1}$ for Si, in good agreement with experiment (~157 W $m^{-1}$ $K^{-1}$, Ref. [2]), and 13.4 W $m^{-1}$ $K^{-1}$ for Al. All the DFT calculations were performed using the Vienna Ab initio Simulation Package (VASP) [3] with the projector-augmented wave (PAW) method [4,5], the recommended PAW potential set, the local-density approximation (LDA), and a plane-wave cutoff of 500 eV. The LDA-optimized lattice constants (5.425 Å for Si, 3.985 Å for Al) were used throughout. The Brillouin zone integration for the primitive cell calculations used the Gaussian smearing with a width of 0.05 eV and 14×14×14 $k$ points for Si, and Methfessel– Paxton broadening with a width of 0.2 eV and 18×18×18 $k$ points for Al.

S4. Workflow for generating TDTR phase data and determining $\kappa_{\mathrm{app}}$

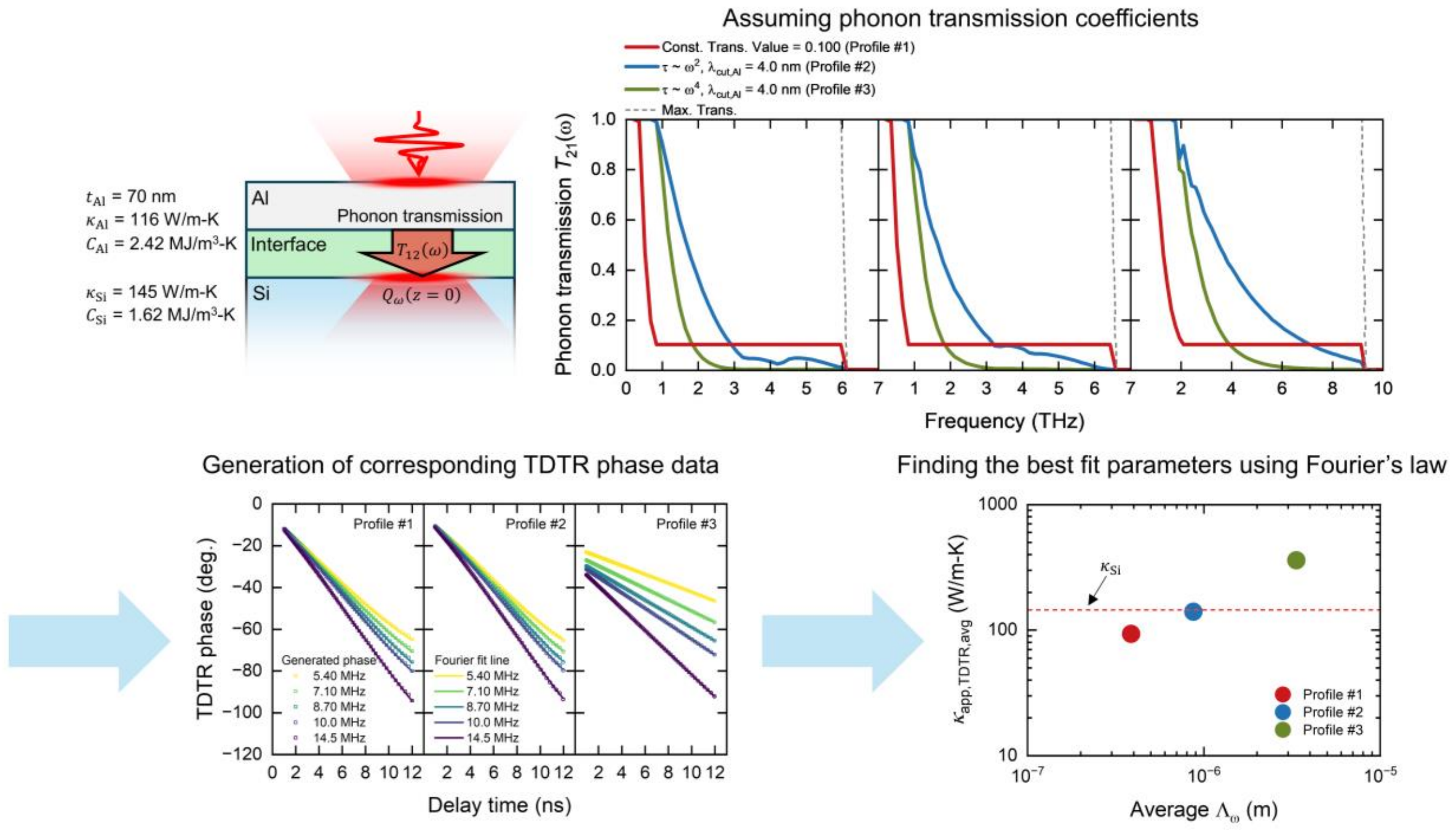

Figure S4. Generation of synthetic TDTR phase data from phonon transmission coefficients and subsequent determination of the apparent substrate thermal conductivity.

The model structure for the TDTR measurement consists of a 70-nm-thick Al film on a Si substrate. The Al layer serves as a transducer that converts the incident laser energy into heat. In the TDTR experiment, a pulsed laser with a repetition rate $f_{\mathrm{rep}}$ = 80.1 MHz is modulated by an electro-optic modulator (EOM) at frequency $f_{\mathrm{mod}}$, so that the heat source oscillates at the frequencies $f_{\mathrm{th}} = |f_{\mathrm{mod}} \pm m f_{\mathrm{rep}}|$, where $m = -\infty \dots \infty$. Phonons transmitted across the Al/Si interface generate an oscillatory surface heat-generation at the top of the Si substrate. The phonon transmission coefficients are determined according to the dominant interfacial scattering mechanisms (details are provided in Sec. S5 and Fig. S5). For each transmission spectrum, the TDTR phase signal is calculated as a function of pump-probe delay time. The resulting synthetic phase data are then fitted with the multilayer Fourier heat-conduction model using two adjustable parameters – the apparent thermal boundary conductance $G_{\mathrm{app}}$ and the apparent substrate thermal conductivity $\kappa_{\mathrm{app}}$ – for each modulation frequency $f_{\mathrm{mod}}$ = 5.40, 7.10, 8.70, 10.0 and 14.5 MHz. Finally, the average $\kappa_{\mathrm{app}}$ is obtained as the mean value over all $f_{\mathrm{mod}}$ results, and the heat-flux-weighted average mean free path $\langle\Lambda_\omega\rangle$ is evaluated using the total of 505 frequencies with $m = -50$ to $50$ at the five different $f_{\mathrm{mod}}$.

S5. Phonon transmission coefficients with extracted $G_{app}$ and $\kappa_{app}$

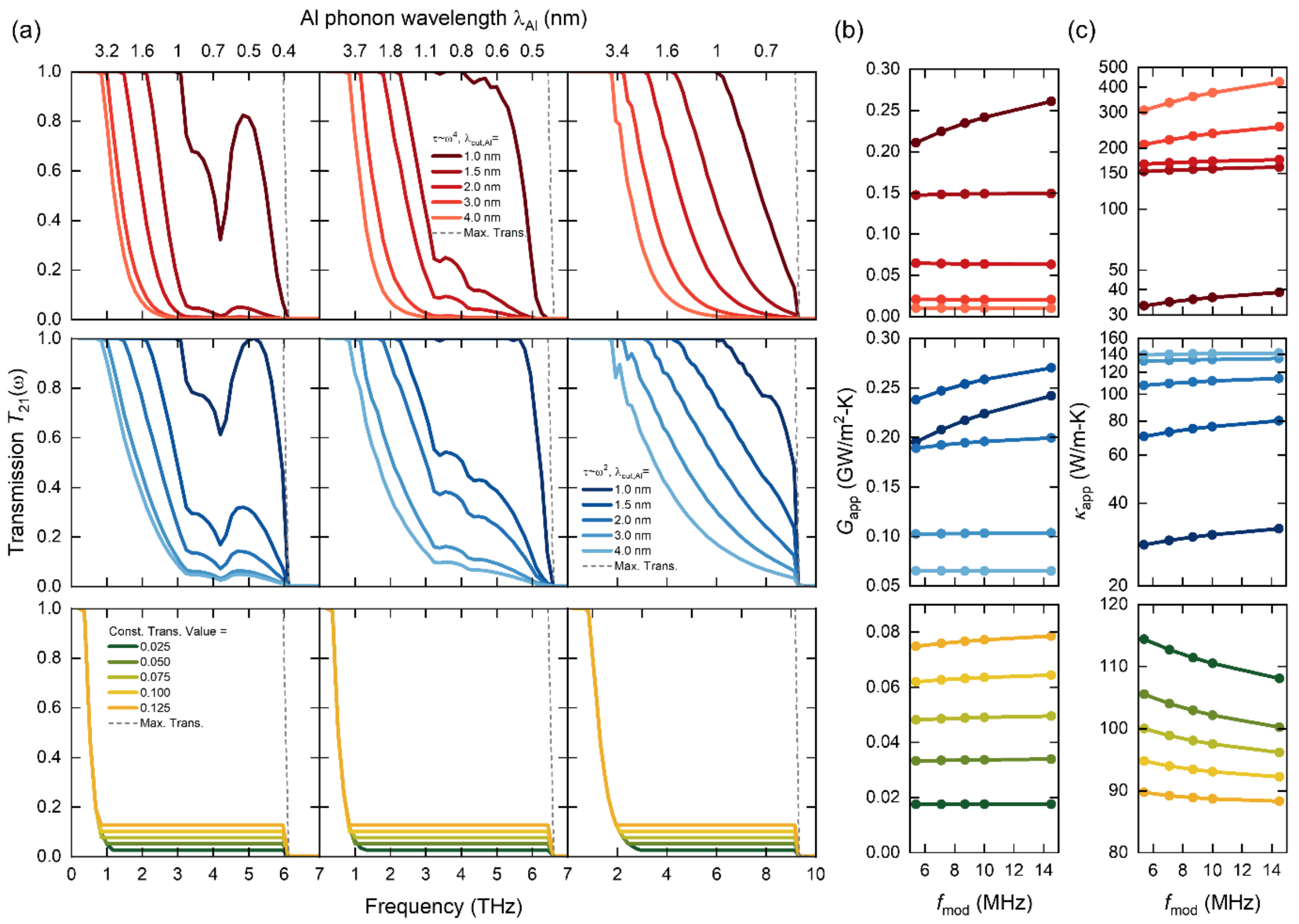

Figure S5. (a) Phonon transmission coefficients $T_{21}(\omega)$ used to generate the TDTR phase data, together with the corresponding extracted values of (b) apparent thermal boundary conductance $G_{app}$ and (c) apparent substrate thermal conductivity $\kappa_{app}$ as a function of the pump modulation frequency $f_{mod}$. Rows from top to bottom correspond to the phonon scattering rate $\omega^n / v^{n-1}$ with $n = 4$ and $n = 2$ , and constant-transmission families described in the text. The dotted line in (a) is the maximum transmission line determined by detailed balance.

The phonon transmission coefficients of Si are plotted as $T_{21}(\omega)$ in Fig. S5(a), so that the low-frequency value reaches the maximum transmission line from detailed balance for Si; the two coefficients remain linked by detailed balance $C_{1\omega} v_{1\omega} T_{12} = C_{2\omega} v_{2\omega} T_{21}$ with medium 1 (Al) and medium 2 (Si). To generate the phonon transmission coefficients, we use an analytic relation between $T_{12}$ and an interface scattering rate $1/\tau_{12}$ for two media labeled 1 (Al) and 2 (Si) [6,7],

$$T_{12} = \frac{\frac{\tau_{12}}{L_z}}{\frac{3}{4v_1} + \frac{1}{2}(1+\beta)\frac{\tau_{12}}{L_z}}.$$

The formula comes from a geometry with recurrent scatterers separated by $L_z$. Here we do not assign a separate value to $L_z$; we treat the ratio $\tau_{12}/L_z$ as the free parameter used to construct $T_{12}$. The factor $\beta = C_1v_1/C_2v_2$ is evaluated mode by mode from the same first-principles inputs as the pBTE. We use a scattering rate $\omega^n/v^{n-1}$. Here, $n$ = 2 and $n$ = 4 correspond to line-defect and point-defect scattering, respectively, and $n$ = 0 represents a broadband mode-conversion layer, such as graphene, inserted at the interface. Furthermore, we include an onset frequency that starts decreasing $T_{21}$ from unity with systematically varied Al phonon wavelength. This describes the variation of the thickness of the interface in a realistic case. Figure S5(b, c) show the apparent thermal boundary conductance $G_{app}$ and apparent substrate thermal conductivity $\kappa_{app}$. They exhibit different functional forms and values under $f_{mod}$ variation for every phonon transmission coefficient case. We attribute this behavior to fitting the actual temperature depth profile across the whole Al/Si structure with the multilayer Fourier heat-conduction model using only two free parameters, $G_{app}$ and $\kappa_{app}$. Therefore, the variation of $G_{app}$ and $\kappa_{app}$ as a function of $f_{mod}$ in TDTR measurement may contain spectral information about the phonon transmission coefficients.